# Prototypes of Using Directional Antenna for Railroad Crossing Safety Applications

Xiaofu Ma[†1], Sayantan Guha[†2], Junsung Choi[†3], Christopher R. Anderson[‡],
Randall Nealy[†4], Jared Withers[⋆], Jeffrey H. Reed[†5], Carl Dietrich[†6]

[†] *Bradley Department of Electrical & Computer Engineering, Virginia Tech, Blacksburg, Virginia, USA*
[xfma[1], sayantg[2], choijs89[3], rnealy[4], reedjh[5], cdietric[6]]@vt.edu

[‡] *Wireless Measurements Group, Department of Electrical & Computer Engineering, U.S. Naval Academy, Annapolis, Maryland, USA*
canderso@usna.edu

[⋆]*Department of Transportation, Federal Railroad Administration, Washington, DC, USA*
jared.withers@dot.gov

*Abstract—* **In this demonstration proposal, we present a prototype of a rapidly deployable and cost-effective railroad crossing early warning system integrated with the railway system. Specifically, the proposed demonstration deal with the safety applications based on dedicated short range communications (DSRC) protocol and devices using our different antennas. We will demonstrate the feasibility and advantages of our proposed system, including the antenna design, system deployment, the over-the-air transmission, and the software applications that we developed for the end users[1].**

## I. INTRODUCTION

Collisions between trains and passenger vehicles at railroad grade crossings are the most common railroad accidents today in the U.S. [1, 2]. Significant vehicle damage, serious passenger injury or even fatalities are the major results of the collisions, because of the size and speed of a train.

Current conventional grade crossing protection systems can be extremely expensive, which is expected to improve in a wireless way. Wireless technologies have been advanced extremely fast in recently years [3], ranging from cellular [4], Wi-Fi [5] to wireless sensor networks [6, 7], etc., and have been applied to various of areas, such as public safety [8], healthcare [9], distributed computing [10], etc. Particularly for transportation, Dedicated Short Range Communication (DSRC) [11], are being explored by the automotive safety industry to provide Intelligent Transportation Services (ITS). This technology can also be adopted for use as a train-to-vehicle collision warning system for unprotected grade crossing [12]. Systems installed on the locomotive would transmit warning messages to the approaching vehicles and roadside warning units. Additionally, such systems could provide the train's driver with information about the potential incoming traffic.

The proposed demonstration will show our DSRC-based warning system using directional antennas. The previously measured results and end user application that we developed will also be demonstrated.

## II. DEMONSTRATION PROTOTYPE

Our proposed demonstration is a prototyping warning system. The communication system consists three major different nodes: an On-Board Unit (OBU) is lodged on the exterior of the train, a Road-Side Unit (RSU) is at the crossing and OBUs are also installed on the vehicles to which the warning messages are directed. Each of these are DSRC units that operate in the 5.9 GHz band. The OBU on the train broadcasts warning messages prior to approaching the crossing, and the warning lasts until the train has cleared the crossing. The warning message is expected to be received at the RSU as well as at the OBU of each of the vehicles near the railway track, thus notifying all vehicles about the incoming train. The RSU acts as a relay for warning messages: it receives the messages from the train and then retransmits them to the vehicles along the road, therefore effectively increasing the coverage range of the warning system. The RSU, thus functions as both a transmitter and a receiver. The overall system concept is shown in Fig. 1.

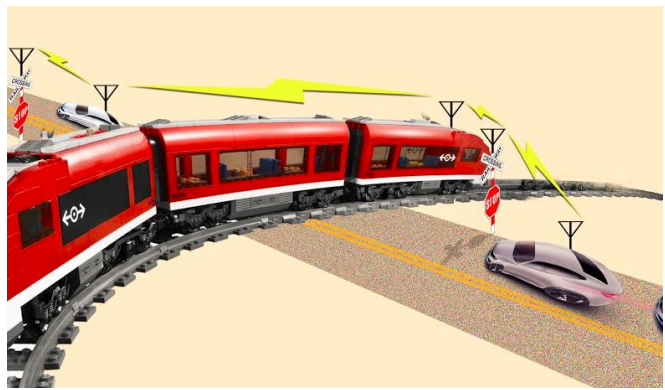

Fig. 1.    The overall system concept.

---

[1] This work was funded in part by the affiliates of Federal Railroad Administration (FRA) and Wireless@Virginia Tech

While vehicle-to-vehicle communications have been well investigated, there are very few studies in the literature that address train warning systems using wireless technologies. Due to the special geographical layout of the crossing of the railroad and road, suitable directional antenna would have advantages for achieving the safety goal. The purpose of our demonstration is to present the feasibility of adopting wireless communication techniques and selecting the suitable antennas for the railroad crossing safety applications.

The system consists of the following four modules:

(1) Antenna modules:

We assemble different types of antennas for directional transmission. The first one is the linear array antenna. It consists of eight Mobile Mark EC012-5900 vertical antennas with a nominal gain of 12 dBi as shown in Fig. 2. The array was fed by an eight way power divider and phase matched cables. The array elements were mounted on a base plate that sets the array spacing. We compare its performance with the omnidirectional antenna, which is a single Mobile Mark EC012-5900 vertical antenna.

In addition, we also demonstrate the over-the-air packet transmission performance of our designed slot array antenna, which is designed based on the practical consideration of the train-to-vehicle communications. Fig. 3 shows the slot array antenna. The array is printed on Rogers RO 4003C microwave substrate with a thickness of 32 mils (0.032 inch). The material is supplied from the manufacturer as a double side copper foil clad panel. Patterns are formed on the board through a photographic process and chemical etching. Vias and connector holes are formed by conventional drilling followed by a chemical metal plating process. The RO 4003C material is compatible with standard circuit board fabrication techniques, so the cost of processing is relatively low. The feed structure is a series of Tee junctions with 71 Ohm quarter wavelength microstrip matching sections and 50 Ohm microstrip transmission lines. A 100 Ohm matching section at each slot element completes the impedance matching to each slot feed point. The slot is shunt fed by a line crossing the slot on the back layer and connected to the ground plane by a via connection. The array feed point is located near the center of the array and is attached to the center conductor of a SMA connector. The shield of the coaxial connector is soldered to the ground plane on the front layer. While this configuration has inferior matching compared to an end launch microstrip to coaxial connection, the end launch method would require a long microstrip line leading to the edge of the board resulting in additional resistive losses.

(2) Communication module:

We use and demonstrate the off-the-shelf commercial DSRC devices for the train-to-vehicle communication together with the different antennas discussed above. The DSRC transmitter power has two types: private level and public safety level. Private power level uses 11 dBm and public safety power level used 23 dBm. Different modulation and coding schemes can be flexibly configured on the devices.

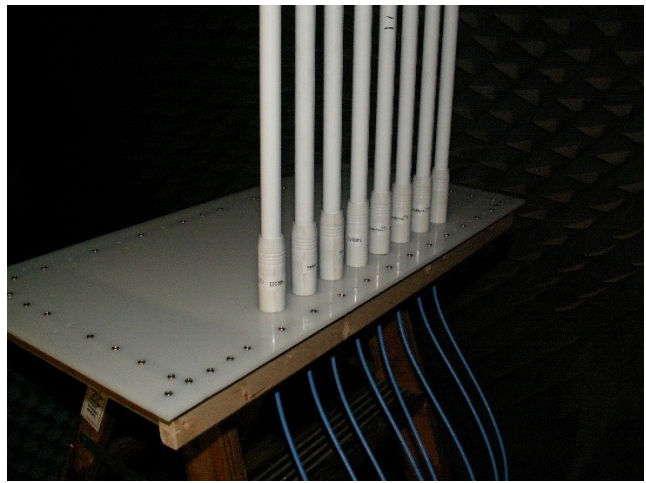

Fig. 2  Linear Array Antenna for Directional Transmission.

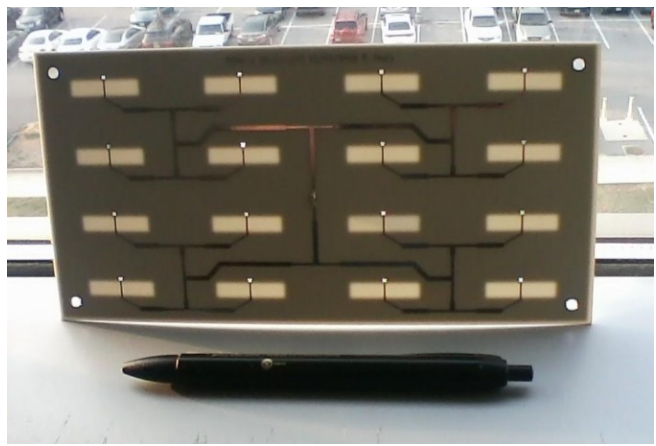

Fig 3.  Slot Array Antenna for Directional Transmission.

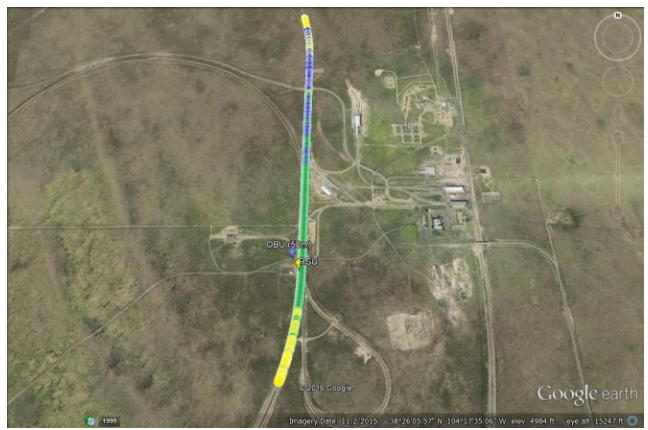

Fig. 4  Google Earth Plot of Received Packets for an Example Case.

(3) User application module:

We develop the application for the end user. Its basic functions include the audio and visual warning messages to the railroad crossing safety application users, such as the vehicle drivers, pedestrians and railroad administrators, etc.

(4) System performance visualization tool:

We develop the performance visualization tool for displaying and analyzing the over-the-air packet transmission and presenting the real-time system performance on Google Earth. One example field measurement results on one track in Transportation Technology Center, Inc., Pueblo, Colorado, U.S. are shown in Fig. 4. It is the case of (a) one transmitter DSRC device that is equipped on one locomotive, (b) one receiver DSRC device as a RSU, and (c) one receiver DSRC device as an OBU that is equipped on an automotive. GPS data based on our measurement was parsed and reformatted to create google earth files. The Google earth plots display the received packets. The northernmost point represents the first point transmitted. The locomotive then continued transmitting as it traveled southward. The white points represent the packets that were broadcast but neither the RSU nor the OBU received the packet. The yellow points represent the packets that only the RSU received. The blue points are the packets that only the OBU received. The green points represent the packets that both the RSU and the OBU received. The yellow and blue place markers are also used to show the locations of the RSU and OBU, respectively.

III. CONCLUSION AND FUTURE WORK

A demonstration of the railroad crossing safety applications are presented. The specially designed directional antenna with DSRC communication system is suitable for train-to-vehicular scenarios. The warning system with end user application can deliver audio and visual warning messages to increase safety level for the drivers. The end user application can be easily deployed on the users' personal mobile devices. For the data analysis of the warning messages, we develop visualization tool that can display the data transmission performance for both OBU and RSU.